\documentclass[preprint,tightenlines,floats,aps,amsmath,amssymb,nofootinbib,prd,showpacs,superscriptaddress]{revtex4}
\usepackage{graphicx} % Include figure files
\usepackage{subfigure}

\usepackage{multirow}
\usepackage{dcolumn} % Align table columns on decimal point
\usepackage{bm}% bold math
\usepackage{hyperref}% add hypertext capabilities
\usepackage{float}
\usepackage{graphicx}
\usepackage{graphics}
\usepackage{epstopdf}
\usepackage{epstopdf}
\usepackage{amsmath}
\usepackage{amssymb}
\usepackage{float}
\graphicspath{{photo/}}

\usepackage{enumerate}

\begin{document}
	\title{ Implications of supermassive neutron star for the form of equation of state of hybrid stars}
	%The Tidal Deformability and Moment of Inertia
	\author{Hongyi Sun}
	\affiliation{School of Physics and Optoelectronics, South China University of Technology, Guangzhou 510641, P.R. China}
	\author{ Dehua Wen\footnote{Corresponding author. wendehua@scut.edu.cn}}
	\affiliation{School of Physics and Optoelectronics, South China University of Technology, Guangzhou 510641, P.R. China}
	\date{\today}
	
	\begin{abstract}
		
		The observations of PSR J0952-0607 ($M=2.35^{+0.17}_{-0.17}~M_{\odot}$) and the second object in GW190814 event ($M=2.59^{+0.08}_{-0.09}~M_{\odot}$) indicate the possible existence of supermassive neutron stars.
		In this work, by using the Constant-Sound-Speed (CSS) parametrization  to describe the equation of state (EOS) of quark matter, the constraints on the EOS parameters from supermassive hybrid stars are investigated through the Maxwell and Gibbs constructions.
		It is shown that to support a supermassive hybrid star (e.g.,  $M=2.5~M_{\odot}$), a lower transition energy density ($\varepsilon_{tran}$), a smaller energy density discontinuity ($\Delta\varepsilon$) and a higher sound speed of quark matter ($c_{sq}$) are favored.   For the constructed hybrid star EOS model, the maximum mass of the corresponding hybrid stars will not meet the lower mass limit of the second object in GW190814 if $\Delta\varepsilon$ takes a value higher than $180~{\rm MeV~fm^{-3}}$.  Moreover, it is confirmed that the supermassive neutron star observation can also rule out the existence of twin stars as a supermassive hybrid star requires a relatively small  $\Delta\varepsilon$.
		Finally, we give a rough estimate of the lower limit of the dimensionless tidal deformability of neutron stars which ranges from 2 to 3.
		
		\begin{flushleft}
		\end{flushleft}
	\end{abstract}
	\maketitle
	
	\section{Introduction}
	
	Neutron star is one of the objects in the universe that have the most extreme environment. As the density of the core is far beyond the reach of terrestrial experiments, neutron stars provide a unique way to the understanding of super dense matter.
	From the surface to the interior, a neutron star is usually  divided into the outer crust, the inner crust, the outer core and the inner core \cite{Yunes2022}. Normally, the equations of state (EOSs) below nuclear saturation density ($\rho_{sat}$) are understood relatively well \cite{Ruster2006, Baldo2007,Chamel2008}. However, for super dense matter, that is, matter in the inner core of neutron stars, hyperons\ \cite{Aguirre2022,Togashi2022,Schaffner-Bielich2002}, free quarks\ \cite{Alford2017,YangS2022,Ivanenko1965}, and even pion-condensation\ \cite{Kampfer1981} may appear, which will soften the high-density EOSs and reduce the maximum mass and radius of neutron stars\ \cite{Haensel2016} (sometimes to a level that cannot meet the observation limit of $2~M_{\odot}$). Overall, the high-density EOSs are still very uncertain.
	Therefore, one has to rely on observations and detailed modeling to learn more about the EOSs of super dense matter.

	Recent observations of the mass-radius measurements from NICER  \cite{Riley2021,Miller2021,Riley2019, Miller2019} have provided important information for improving the understanding of neutron stars and  the EOSs of dense nuclear matter\ \cite{Miao2022,Li2021}.
	Besides, the detection of the binary neutron star merger, GW170817 \cite{Abbott2017}, has constrained the tidal deformability of $1.4 ~M_{\odot}$ neutron star to $\Lambda_{1.4}= 190^{+390}_{-120}$, which in turn has been employed to further constrain the pressure at 2$\rho_{sat}$\ \cite{Abbott2017,Abbott2018}.
	These neutron star properties have also been theoretically calculated using a wide range of EOSs (moment of inertia \cite{I-Love-C,Jiang2020APJ,Silva2021PRL,YuxiLi2022CQG,syang2022PRD}, radius \cite{Abbott2018,De2018,Malik2018,Silva2021PRL}, the parameter of f-mode oscillation \cite{Wen2019PRC,Pratten2020}  and gravitational binding energy \cite{JiangRR2019PRD,SunWJ2020PRD}).
	In addition, the discovery of supermassive compact stars is likely to reshape our understanding of the maximum mass of neutron stars\ \cite{Romani2020,Abbott2020}.
	In the GW190814 event a merger of a black hole with an unknown compact object with mass of $M=2.59^{+0.08}_{-0.09}~M_{\odot}$  was observed\ \cite{Abbott2020}, and the heaviest known galactic neutron star PSR J0952-0607 has a mass of $M=2.35^{+0.17}_{-0.17}~M_{\odot}$\ \cite{Romani2020}.
	In light of such discoveries, the study of supermassive compact stars can help to better understand the possibility of the phase transition and the appearance of quark phase in compact stars\ \cite{Dexheimer2021,Dexheimer2021T}.

	It has long been noted that with the appearance of free quark matter, a new class of compact stars composed of  a deconfined quark core and a hadronic outer shell may arise\ \cite{Alford2013,Alford2016,Blaschke2018}. For better distinction, compact stars composed of  pure hadronic matter are  usually called as normal neutron stars, and the stars with a deconfined quark core and a hadronic outer shell are referred to as  hybrid stars\ \cite{Zacchi2016,Alford2013,Alford2016,Glendenning2000,Glendenning2000M,Blaschke2018,Blaschke2019,Li2020}.
	Many previous works have explored the possibility of the existence of supermassive hybrid stars with mass close to the mass of the second object in GW190814 and investigated the macroscopic and internal microscopic properties of such supermassive hybrid stars\ \cite{Blaschke2021,Christian2021,Drischler2021,Bozzola2019,Zhang2019}.
	However, since the GW170817 event favors the soft EOSs, and the existence of supermassive hybrid stars puts forward a requirement for the stiff EOSs, it is important to resolve this conflict and use the GW170817 event to constrain the properties of supermassive stars\ \cite{Lim2021,Godzieba2021,Tews2021,Tan2020,Tsokaros2020,Horvath2021}.

	In this work, we will use the Maxwell\ \cite{Christian2022,Ranea-Sandoval2016} and Gibbs constructions\ \cite{Blaschke2019,Macher2005} to connect hadronic matter and quark matter, and describe quark matter by the CSS parametrization\ \cite{Alford2013}.
	Other quark models, such as NJL\ \cite{Zuo2022,Benic2015} and MIT bag\ \cite{Johnson1976} models, may be more realistic.
	For example, Beni\'{c} $et~al.$ investigated the high-mass twin-stars by using  NJL model with higher-order quark interactions\ \cite{Benic2015}.
	However, the CSS parametrization can fit these quark models well in some conditions\ \cite{Contrera2022,Cierniak2021}, and its parameters are more adjustable.
	So this ansatz provides us with the opportunity to change the parameters and apply the mass ranges of PSR J0952-0607 and the second object in GW190814 through a more generalized way\ \cite{Christian2021}.
	Further, we will discuss the mass-radius relations and tidal deformability of hybrid stars\ \cite{Tsaloukidis2022,Montana2019,Li2021}, and then investigate the constraint of the $2.5~M_{\odot}$ (which conforms to the mass range of PSR J0952-0607 and GW190814 event)  hybrid star properties\ \cite{Tan2020,Dexheimer2021,Sen2022,Cierniak2022} on the parameters of EOSs.
	It is worth pointing out that supermassive neutron stars may also appear in other forms, such as pure
	nucleonic neutron stars\ \cite{Huang2020,Kanakis-Pegios2021}, quark stars\ \cite{Albino201,Sen2022B,Roupas2021}, fast-spinning neutron stars\ \cite{Wang2022,Dexheimer2021T} and dark matter admixed neutron stars\ \cite{Lee2021}.

	This paper is organized as follows. In Sec.II, the basic formulas for the macroscopic properties of neutron stars are briefly introduced.
	The construction of hybrid star EOSs through the CSS parameterization is presented in Sec.III.
	In Sec.IV, the mass-radius relations of hybrid stars are presented and discussed.
	The tidal deformability of $2.5~M_{\odot}$ hybrid stars are also investigated.
	Finally, a summary will be given in Sec.V.

	\section{Basic formulas for the macroscopic properties of neutron stars}
	
	Inputting the EOSs of dense nuclear matter, the mass and radius of a neutron star can be calculated through the Tolman-Oppenheimer-Volkoff (TOV) equation\ \cite{Tolman1939,Oppenheimer1939}
	\begin{equation}
		\frac{dp}{dr}=-\frac{G\varepsilon(r)M(r)}{c^{2}r^{2}}\left(1+\frac{p(r)}{\varepsilon(r)}\right)\left(1+\frac{4\pi r^{3}p(r)}{M(r)c^{2}}\right)\left(1-\frac{2GM(r)}{c^{2}r}\right)^{-1},
	\end{equation}
	supplemented with the differential equation of mass
	\begin{equation}
		\frac{dM}{dr}=\frac{4\pi r^{2}\varepsilon(r)}{c^{2}},
	\end{equation}
	where $M(r)$, $p(r)$ and $\varepsilon(r)$ are the enclosed mass, pressure and energy density at the radius $r$.
	
	We solve an additional equation for the tidal deformability of neutron stars, defined as
	\begin{equation}
		\lambda=\frac{2}{3}R^{5}k_2.
	\end{equation}
	The second (quadrupole) tidal Love number $k_2$ can be solved by the following equation  \cite{Hinderer2010,Kanakis-Pegios2020}
	
	\begin{equation}
		\begin{aligned}
			k_{2}=& \frac{8}{5}x^{5}(1-2x)^{2}(2-y_{R}+2x(y_{R}-1))
			(2x(6-3y_{R}+3x(5y_{R}-8))\\
			& +4x^{3}(13-11y_{R}+x(3y_{R}-2)+2x^{2}(1+y_{R}))\\
			& +3(1-2x)^{2}(2-y_{R}+2x(y_{R}-1))ln(1-2x))^{-1},
		\end{aligned}
	\end{equation}
	where $x=GM/Rc^{2}$ is the compactness of neutron stars, and  $y_{R}$ is determined by solving the following differential equation
	\begin{equation}
			r\frac{dy(r)}{dr}+y^2(r)+y(r)F(r)+r^2Q(r)=0
		\label{Eq.1}
	\end{equation}
	with $F(r)$ and $Q(r)$ being the functions of $M(r)$, $p(r)$ and $\varepsilon(r)$\ \cite{Kanakis-Pegios2020}
	\begin{equation}
		F(r)=\left[1-\frac{4\pi r^{2}G}{c^4}(\varepsilon(r)-p(r))\right]\left(1-\frac{2M(r)G}{rc^2}\right)^{-1},
	\end{equation}
	\begin{equation}
	\begin{aligned}
		r^2Q(r)=& \frac{4\pi r^{2}G}{c^4}\left[5\varepsilon(r)+9p(r)+\frac{\varepsilon(r)+p(r)}{\partial p(r)/\partial\varepsilon(r)}\right]\\
		&\times\left(1-\frac{2M(r)G}{rc^2}\right)^{-1}-6\left(1-\frac{2M(r)G}{rc^2}\right)^{-1}\\
		&-\frac{4M^{2}(r)G^{2}}{r^2c^4}\left(1+\frac{4\pi r^{3}p(r)}{M(r)c^2}\right)^{2}\left(1-\frac{2M(r) G}{rc^2}\right)^{-2}.
	\end{aligned}
\end{equation}
Once $y(r)=y_R$ at the radius $R$ of neutron stars is provided,
 the dimensionless tidal deformability parameter $\Lambda=\lambda({GM/c^{2}})^{-5}$ can be obtained.

		\section{EOS of Hybrid star}
		In this section we introduce how to construct the hybrid star EOSs, where a  phase transition from hadronic matter to quark matter is considered.
		\begin{figure}[h]
			{\centering
				\includegraphics[width=0.5\textwidth]{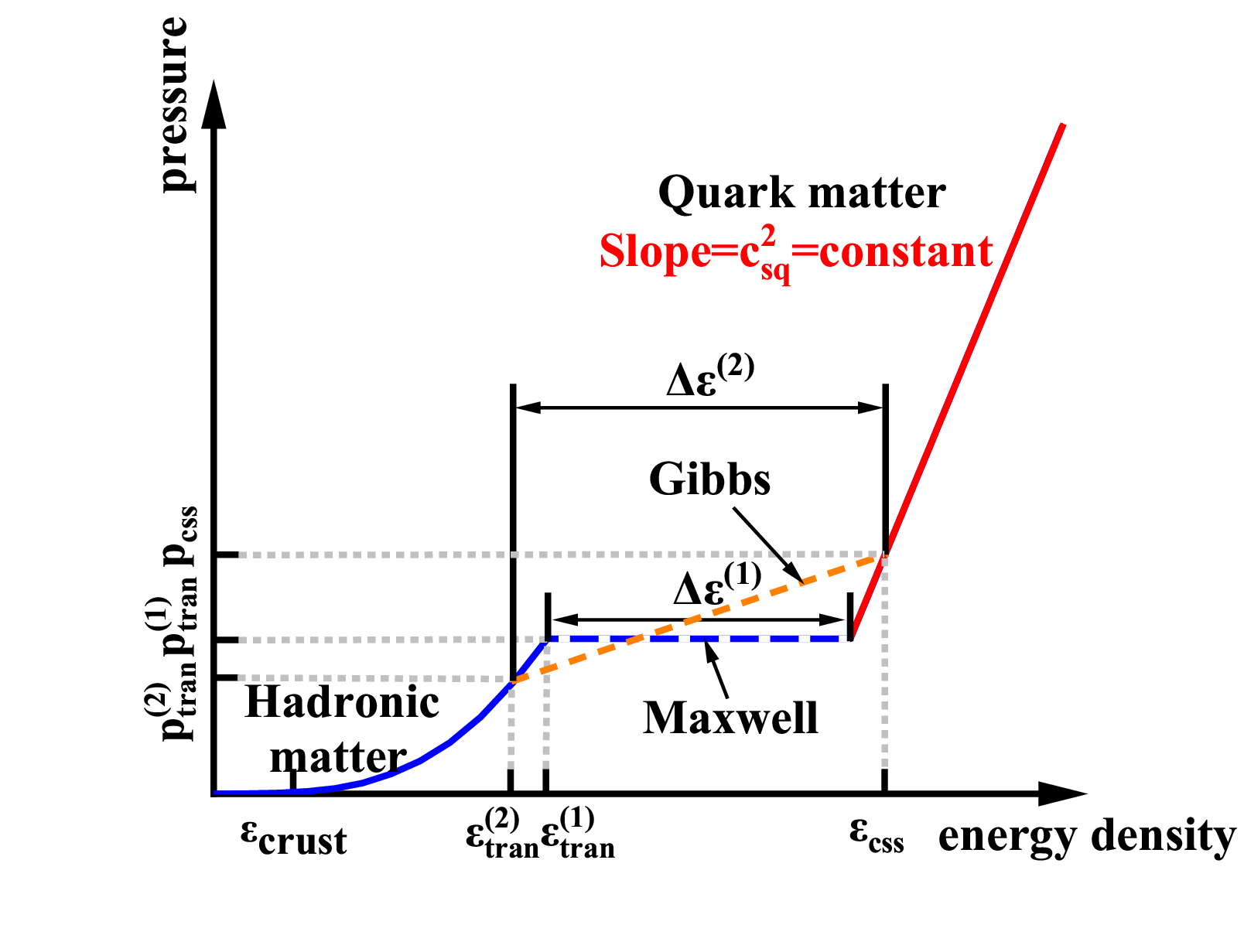}
				\caption{ The schematic form of the hybrid star EOSs with the CSS parametrization and Maxwell/Gibbs construction.}
				\label{Fig.1}}
		\end{figure}
		For the outer crust and the inner crust, the BPS EOS\ \cite{Baym1971} and NV EOS \ \cite{Negele1973} are adopted, respectively.
		In the outer core where the energy density of hadronic matter is lower than the transition energy density $\varepsilon_{tran}$, two hadronic EOSs with different stiffness, the APR3 EOS (soft)\ \cite{APR4} and DDME2 EOS (stiff)\ \cite{DDME2} are employed in order to explore how the stiffness of hadronic matter effects the properties of hybrid stars.
		In the inner core, after the  phase transition (symbolized by the energy density discontinuity $\Delta\varepsilon$), quark  matter with energy density higher than $\Delta\varepsilon+\varepsilon_{tran}$ is described by the CSS parametrization\ \cite{Alford2013}.
		
		For the Maxwell construction\ \cite{Glendenning1996}, the formula form of hybrid star EOSs is given by
		\begin{equation}
			\label{eq:H1}
			\varepsilon(p)=\left\{
			\begin{aligned}
				& \varepsilon_{h}(p), & p\leqslant p_{tran} \\
				& \varepsilon_{tran}+\Delta\varepsilon+c_{sq}^{-2}(p-p_{tran}), & p>p_{tran},
			\end{aligned}
			\right.
		\end{equation}
		where $\varepsilon_{h}(p)$ represents the hadronic EOSs, $c_{sq}^{2}$ is the squared sound speed of quark matter (in the following, $c_{sq}^{2}$ is in unit of $c^{2}$), and $\varepsilon_{tran}$ $(p_{tran})$ is the transition energy density (transition pressure).
		For the Gibbs construction, the phase transition is described by a polytrope EOS $p(\rho)=K_{m}\rho^{\Gamma_{m}}$\ \cite{Montana2019,Tsaloukidis2022} to account for a mixed phase of hadron and quark. The hybrid star EOSs with the Gibbs construction are given by
		\begin{equation}
			\label{eq:H2}
			\varepsilon(p)=\left\{
			\begin{aligned}
				& \varepsilon_{h}(p), & p\leqslant p_{tran} \\
				& A_{m}(p/K_{m})^{1/\Gamma_{m}}+p/(\Gamma_{m}-1), & p_{tran}< p\leqslant p_{css} \\
				& \varepsilon_{css}+c_{sq}^{-2}(p-p_{css}), & p>p_{tran},
			\end{aligned}
			\right.
		\end{equation}
		with $\Gamma_{m}=1.03$. This choice allows us to get a strong enough softening of the EOS, which is significantly different from the Maxwell construction.
		The values of $K_{m}$ and $A_{m}$ are obtained by requiring that $p$ and $\varepsilon$ are continuous at the transition point. $\varepsilon_{css}$ and $p_{css}$ is the energy density and pressure at the begin of quark phase.
		It is noted that the energy density discontinuity $\Delta\varepsilon$ is not defined in Eq \ref{eq:H2}. In this case, we assign the increment of $\varepsilon(p)$  during the mixed phase to its value.
		The hybrid star EOSs are also illustrated in Fig.\ \ref{Fig.1}.

		\section{Constraint on EOS through Macroscopic Properties of Supermassive Hybrid Star}

		\subsection{Constraint through Mass-Radius Relations of Hybrid Stars}
		
\begin{figure}[h]
			{\centering
				\includegraphics[width=1\textwidth]{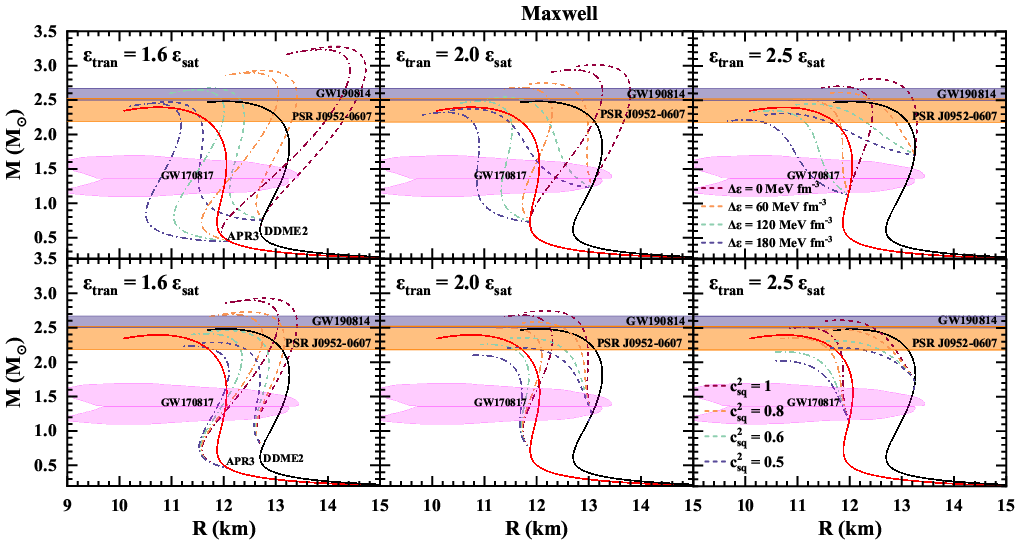}
				\caption{The effect of energy density discontinuity $\Delta\varepsilon$ (upper panels, where $\Delta\varepsilon=0, 60, 120, 180 ~{\rm MeV~fm^{-3}}$) and the squared sound speed of quark matter $c_{sq}^{2}$ (lower panels, where $c_{sq}^{2}=1, 0.8, 0.6, 0.5$) on the M-R relations for hybrid stars with the Maxwell construction, where  $c_{sq}^{2}$ and $\Delta\varepsilon$ are fixed as $c_{sq}^{2}=1$ and $\Delta\varepsilon=60~ {\rm MeV~fm^{-3}}$ for the upper panels and lower panels, respectively.
					The transition energy density $\varepsilon_{tran}$ adopts $1.6 ~\varepsilon_{sat}$ (left), $2.0~ \varepsilon_{sat}$ (middle), $2.5~ \varepsilon_{sat}$ (right).
					The dash (dot dash) lines are the M-R relations of hybrid stars with hadronic matter described by the DDME2 (APR3) EOS.
					The black (red) solid line is the M-R relation of normal neutron stars calculated by the DDME2 (APR3) EOS.
					The orange and purple shaded areas show the mass range of PSR J0952-0607\ \cite{Romani2020} and the second object in GW190814\ \cite{Abbott2020}, while the magenta shaded areas are the M-R constraint from GW170817\ \cite{Abbott2018}.
				}
				\label{Fig.2}}
		\end{figure}

\begin{figure}[h]
			{\centering
				\includegraphics[width=1\textwidth]{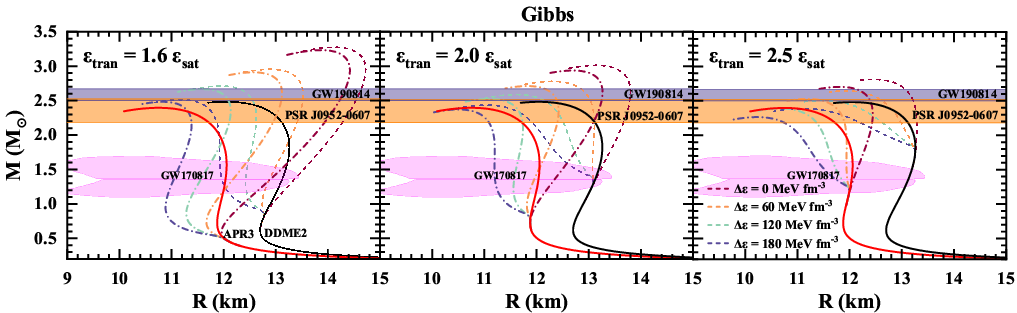}
				\caption{The effect of energy density discontinuity $\Delta\varepsilon$ on the M-R relations for hybrid stars with the Gibbs construction. The parameters are the same as in the upper panels of Fig.\ \ref{Fig.2}.
					The observational constraints are also plotted.
				}
				\label{Fig.3}}
		\end{figure}

		As we know, based on the TOV equation, each EOS corresponds to a unique mass-radius (M-R) relation.
		After constructing the hybrid star EOSs by using the APR3 and DDME2  to describe hadronic matter, the effects of energy density discontinuity $\Delta\varepsilon$ and sound speed $c_{sq}$ on the M-R relations of hybrid stars at different transition energy density $\varepsilon_{tran}$ are investigated, and the results of Maxwell construction are shown in Fig.\ \ref{Fig.2},
where the squared sound speed is fixed as $c_{sq}^{2}=1$ in the upper panels, and the energy density discontinuity is fixed as $\Delta\varepsilon=60~{\rm MeV~fm^{-3}}$ in the bottom panels.
The effect of $\Delta\varepsilon$ on the M-R relations of hybrid stars are also shown for the Gibbs construction in Fig.\ \ref{Fig.3}.
		The M-R relations of normal neutron stars (without the phase transition) are also displayed in Fig.\ \ref{Fig.2} and Fig.\ \ref{Fig.3} as comparison.
		Here we adopt the APR3 representing the relatively soft EOS and DDME2 representing the relatively stiff EOS to describe the hadronic part in neutron stars. It is shown that the maximum mass of normal neutron stars supported  by the APR3/DDME2 EOS can be up to near $2.4$/$2.5~M_{\odot}$.

		It can be seen from Fig.\ \ref{Fig.2} and Fig.\ \ref{Fig.3} that both of the M-R relations of normal neutron stars and hybrid stars meet the constraint of GW170817.
		For the upper three panels of Fig.\ \ref{Fig.2}, $c_{sq}^{2}=1$ means that the stiffest EOS is adopted in quark matter, then the corresponding M-R relation gives the upper limit of stellar mass under a given set of $\Delta\varepsilon$ and $\varepsilon_{tran}$.
		The upper panels show that the increase of $\Delta\varepsilon$ will lead to a reduction of the maximum mass of hybrid stars with the Maxwell construction. When the $\Delta\varepsilon$ takes a relatively high value (e.g., $\Delta\varepsilon=180~{\rm MeV~fm^{-3}}$), the maximum mass of the corresponding hybrid stars will not meet the lower mass limit of the second object in GW190814.
		For the Gibbs construction (see Fig.\ \ref{Fig.3}), a higher energy density discontinuity ($\Delta\varepsilon>180~{\rm MeV~fm^{-3}}$) will induce a similar result. This means that if observations confirm the existence of supermassive neutron stars, it will placed a well constraint on the upper limit of $\Delta\varepsilon$  for EOS model with the quark phase transition.
		In addition, the increase of $\Delta\varepsilon$ will also lead to a reduction of the radius, which is consistent with the results of Refs. \ \cite{Christian2018,Li2021,Deloudis2021}.
		However, compared to the Maxwell construction, the Gibbs construction will result in  a less reduction of radius under  same $\Delta\varepsilon$.
		As for the influence of transition energy density $\varepsilon_{tran}$, we can see that with the increase of $\varepsilon_{tran}$, the maximum mass of hybrid stars  decreases.

		As we know, due to the  phase transition caused by the appearance of quark matter in the core of neutron stars, a new branch with the higher central density than that of the normal neutron stars may appear.
		The new branch alone can lead to the appearance of twin stars with same mass but different radii due to different internal constitutions\ \cite{Blaschke2019,Li2021,Li2022}. From Fig.\ \ref{Fig.2} and Fig.\ \ref{Fig.3}, we can see that the M-R relations do not show the presence of twin stars.
		As discussed in Ref. \cite{Blaschke2018}, the following three quantities, including the transition energy density $\varepsilon_{tran}$, the sound speed $c_{sq}$  and the energy density discontinuity $\Delta\varepsilon$, are the key factors to determine the existence of twin stars. Here, it is the weak $\Delta\varepsilon$ that causes the miss of twin star phenomenon\ \cite{Christian2021}.
		The small $\Delta\varepsilon$ also leads to the intersection of the M-R curves of normal neutron stars and  hybrid stars, which means that it is difficult to determine the existence of the phase transition only through the M-R relations.

		From the lower three panels in Fig.\ \ref{Fig.2}, we can see that for hybrid star EOSs with  energy density discontinuity fixed as $\Delta\varepsilon=60~{\rm MeV~fm^{-3}}$ and with  transition energy density $\varepsilon_{tran}$ higher than $1.6 ~\varepsilon_{sat}$, the corresponding hybrid stars will not meet the lower mass limit of the second object in GW190814 as $c_{sq}^{2}$ reduces to 0.6.
		When considering the Gibbs construction, it leads to the same result.
	    This means that if a supermassive neutron star is confirmed in the future, it will constrain the lower limit of quark matter‘s sound speed.

		\subsection{Constraint through Tidal Deformability  of Supermassive Hybrid Stars ($M~=~$2.5$~M_{\odot}$)}
		\begin{figure*}[h]
			\centering
			\includegraphics[width=1\textwidth]{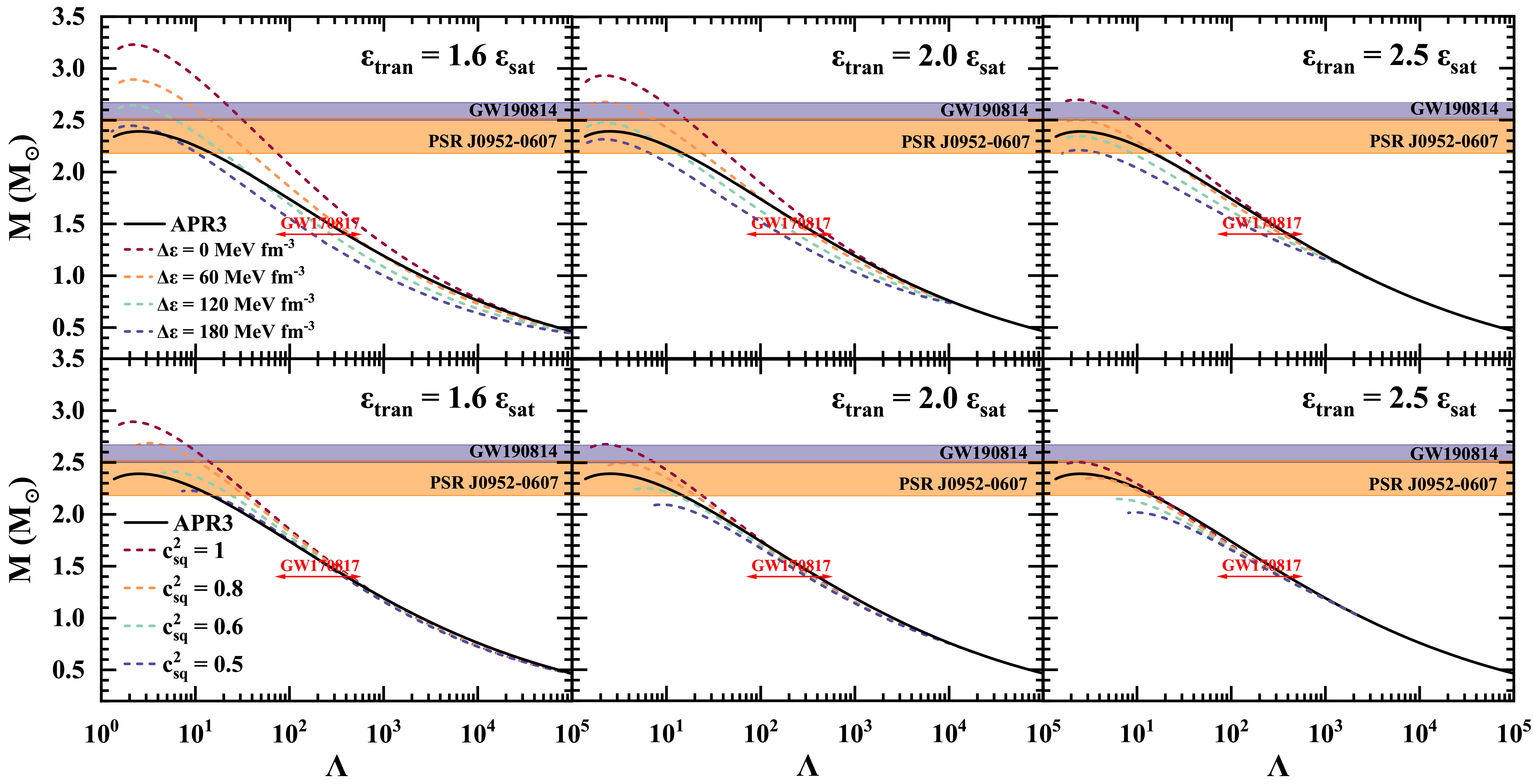}
			\caption{The  mass-tidal deformability (M-$\Lambda$) relations of hybrid stars with the Maxwell construction at different energy density discontinuity $\Delta\varepsilon$ (upper  panels) and squared sound speed of quark matter $c_{sq}^{2}$ (lower  panels),  where the $c_{sq}^{2}$ and $\Delta\varepsilon$ are fixed as $c_{sq}^{2}=1$ and $\Delta\varepsilon=60~ {\rm MeV~fm^{-3}}$ for the upper panels and lower panels, respectively.
				The $\varepsilon_{tran}$ adopts the same values as in Fig.\ \ref{Fig.2}.
				For simplicity,  only the APR3 EOS is considered  here, and the black solid line is the M-$\Lambda$ relation of the corresponding normal neutron stars.
				The constraints from PSR J0952-0607\ \cite{Romani2020}, the second object in GW190814\ \cite{Abbott2020} and GW170817\ \cite{Abbott2018} are also plotted.
			}
			\label{Fig.4}
		\end{figure*}
		
		The influence of the energy density discontinuity $\Delta\varepsilon$ (upper panels) and sound speed $c_{sq}$ (lower panels) on the tidal deformability of hybrid stars with the Maxwell construction are shown in Fig.\ \ref{Fig.4}.
		As shown in the upper panels, at a relatively low transition energy density (e.g., $\varepsilon_{tran}=1.6 ~\varepsilon_{sat} $),  the decrease of $\Delta\varepsilon$ will obviously increase the tidal deformability, and when the $\Delta\varepsilon$ approaches zero with the squared sound speed of quark matter fixed at $c_{sq}^{2}=1$, the tidal deformability will exceed the constraint of GW170817.
		This result indicates that the observation of GW170817 will place some constraints on the upper limit of  $c_{sq}$ and the lower limit of $\Delta\varepsilon$ and $\varepsilon_{tran}$.
		The tidal deformability of hybrid stars with the Gibbs construction is slightly greater than that with the Maxwell construction, which can also be inferred from their different impacts on the radius. However, due to the small difference in tidal deformability caused by the two phase transition structures, the observation of GW170817 will impose the similar constraints on the parameters of hybrid star EOSs with the Gibbs construction.

		From the upper panels ($c_{sq}^{2}=1$), it is interesting to note that even though different M-$\Lambda$ curves correspond to  different maximum masses as the $\Delta\varepsilon$ changes, the maximum mass hybrid stars all correspond to the similar values of tidal deformability.
		Moreover, the lower panels show that the decrease of $c_{sq}^{2}$ will increase the tidal deformability of the maximum mass hybrid stars, while the $\varepsilon_{tran}$ does not have a significant impact on the tidal deformability of the maximum mass hybrid stars.
		In addition, although  Fig.\ \ref{Fig.4} only shows the results for APR3 EOS with the Maxwell construction, the calculations for  DDME2 EOS and Gibbs construction give the similar results.
		So the results in the upper panels of Fig.\ \ref{Fig.4} provide a useful clue to the lower limit of the tidal deformability for the maximum mass hybrid stars.
		A rough estimate of the lower limit of the dimensionless tidal deformability for the maximum mass hybrid stars ranges from 2 to 3.
		In fact, this can also be seen as the lower limit for all neutron stars since the tidal deformability of the maximum mass neutron stars is the smallest in the neutron star sequences.

			\begin{figure*}[h]
			\centering
			\includegraphics[width=1\textwidth]{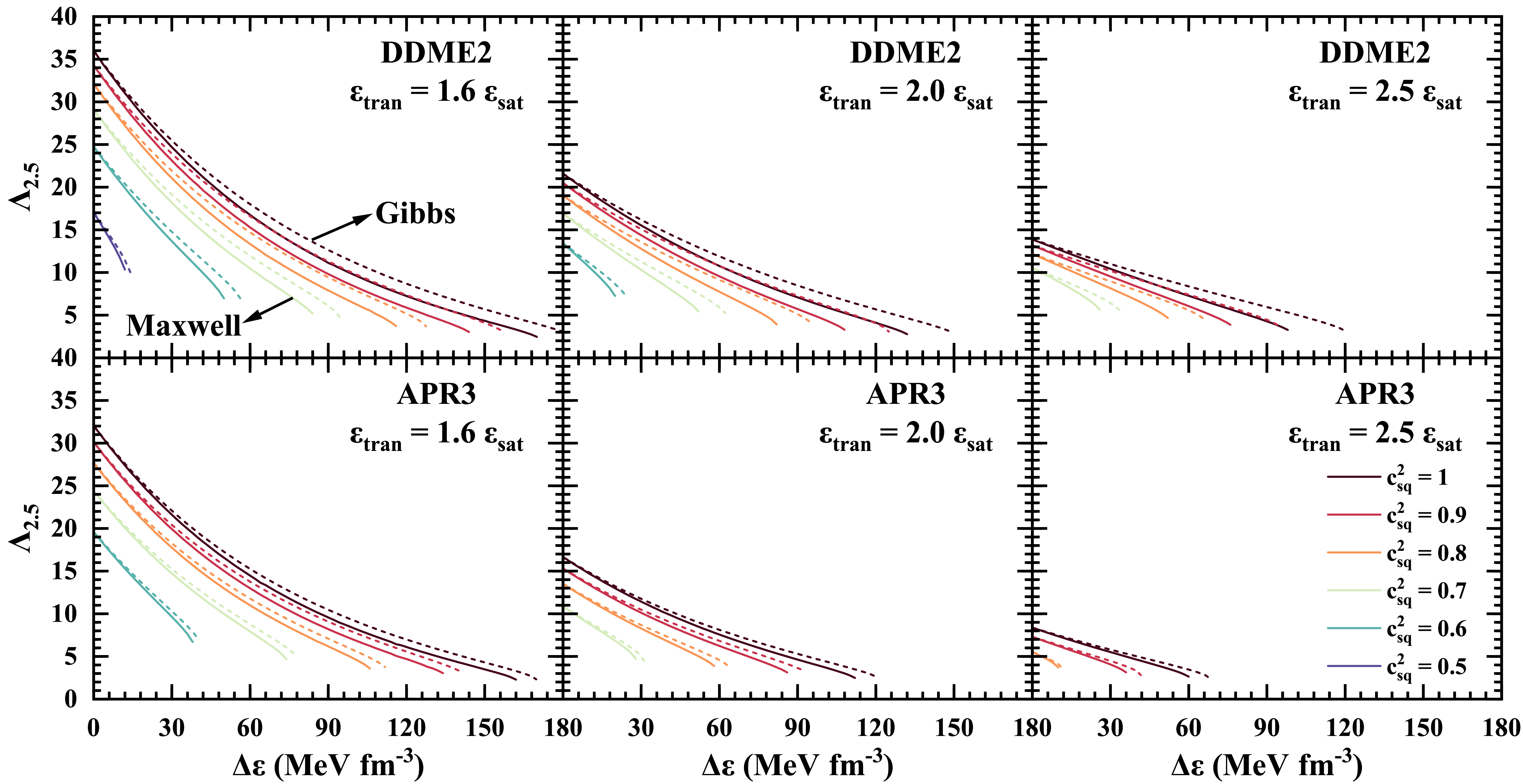}
			\caption{
				The tidal deformability as a function of energy density discontinuity $\Delta\varepsilon$ at different $c_{sq}^{2}$ ($c_{sq}^{2}=1, 0.9, 0.8, 0.7, 0.6, 0.5$), where the stellar mass is fixed as 2.5 $M_{\odot}$.
				The solid and dash lines represent the results of Maxwell and Gibbs constructions, respectively.
				$\varepsilon_{tran}$ adopts the same values as in Fig.\ \ref{Fig.2}.
				The DDME2 (APR3) EOS is taken to describe hadronic matter in the upper (bottom) panels.
			}
			\label{Fig.6}
		\end{figure*}
		
	In order to illustrate how the tidal deformability of supermassive hybrid stars is affected by the relative parameters, the tidal deformability $\Lambda$ as a function of the energy density discontinuity $\Delta\varepsilon$ at different $\varepsilon_{tran}$ and $c_{sq}^{2}$ for the DDME2 EOS model (upper  panels) and  APR3 EOS model (lower panels) is drawn in Fig.\ \ref{Fig.6}, where the stellar mass is fixed as 2.5 $M_{\odot}$.
It is easy to see that for a hybrid star with fixed mass, its tidal deformability with different phase transition structures will both decrease with the decrease of sound speed $c_{sq}$  and with the increase of transition energy density $\varepsilon_{tran}$ or energy density discontinuity $\Delta\varepsilon$, which can be understood by the M-R relations in Fig.\ \ref{Fig.2} and Fig.\ \ref{Fig.3}, that is,   a smaller $c_{sq}$, a higher $\varepsilon_{tran}$, or a larger $\Delta\varepsilon$ corresponds to a smaller stellar radius for supermassive hybrid stars and further gives a smaller tidal deformability.
It is worth noting that for the same energy density discontinuity, the tidal deformability of 2.5 $M_{\odot}$ hybrid star with the Gibbs construction is greater (i.e., the dash lines are higher than the solid lines in Fig.\ \ref{Fig.6}), which is due to the fact that the hybrid star EOSs with the Gibbs construction are stiffer than that with the Maxwell construction under the same parameters.

		Comparing the upper and lower graphes of Fig.\ \ref{Fig.6}, one can see that if the observed dimensionless tidal deformability of  2.5 $M_{\odot}$ neutron star is at around $\Lambda_{2.5}\sim35$, which can be detected under the precision ($\sigma_{\tilde{\Lambda}}<20$) of Einstein telescope\ \cite{Punturo2010,Chatziioannou2022}, then for both the phase transition structures, a smaller tidal deformability of 2.5 $M_{\odot}$ neutron star (e.g., $\Lambda_{2.5}\lesssim15$) will be excluded in an ideal scenario, and only a smaller energy density discontinuity ($\Delta\varepsilon< 75~\textrm{MeV fm} ^{-3}$, which becomes smaller for a higher $\varepsilon_{tran}$  and for the Maxwell construction), a lower transition energy density ($\varepsilon_{tran}<2.5 ~\varepsilon_{sat}$) and a higher  squared sound speed ($c_{sq}^{2}>0.5$) are possible in hybrid stars.
		Conversely, for a small tidal deformability, such as $\Lambda_{2.5}\sim10$, even the third-generation gravitational detectors cannot reach the required precision, so it is hard to make effective constraints on the EOS parameters when the tidal deformability of supermassive neutron stars is small.
		%Therefore, the observation of such supermassive neutron stars requires more accurate gravitational wave detection instruments in the far future. For example, if $\Lambda_{2.5}\sim10$ is observed with $\Delta\Lambda<5$, then $\Delta\varepsilon$ should take a relatively large value, and a higher $\varepsilon_{tran}$ (e.g. $\varepsilon_{tran}>2.5 ~\varepsilon_{sat}$) is allowed.

		In  Fig.\ \ref{Fig.6}, it  also can directly show the ranges of parameters $\Delta\varepsilon$, $c_{sq}$ and $\varepsilon_{tran}$ that can support  2.5 $M_{\odot}$ hybrid star. For example, a higher transition energy density requires a higher sound speed  and a smaller energy density discontinuity. In addition, a softer EOS (e.g., APR3) will narrow the ranges of parameters in order to support 2.5 $M_{\odot}$ hybrid star.

		\section{summary}
		
		The observations of PSR J0952-0607 ($M=2.35^{+0.17}_{-0.17}~M_{\odot}$) and the second object in GW190814 event ($M=2.59^{+0.08}_{-0.09}~M_{\odot}$) indicate the possible existence of supermassive neutron stars.
		The possibility of the existence of supermassive neutron stars provides an opportunity to understand the high density matter from a new angle.
		In this work,  the Constant-Sound-Speed (CSS) parametrization is adopted to describe the EOS of quark matter for the hybrid star EOS, and the Maxwell/Gibbs construction is used to connect hadronic and quark phase.
		By using the constructed EOSs, the constraints on the EOS parameters through the properties of supermasive hybrid stars are probed.
		The main conclusions of this work are summarized as follows.
		\begin{enumerate}[(1)]
			\item	
			It is shown that to support a supermassive hybrid star (e.g.,  $M=2.5~M_{\odot}$), a lower transition energy density ($\varepsilon_{tran}$), a smaller energy density discontinuity ($\Delta\varepsilon$) and a higher sound speed of quark matter ($c_{sq}$) are favored.   For the constructed hybrid star EOSs model with the Maxwell and Gibbs constructions, the maximum mass of the corresponding hybrid stars will not meet the lower mass limit of the second object in GW190814 if the $\Delta\varepsilon$ takes a value higher than $180~{\rm MeV~fm^{-3}}$. If observations confirm the existence of supermassive neutron stars, it will place a well constraint on the upper limit of $\Delta\varepsilon$  for the hybrid star EOSs.
			\item
			As  supermassive hybrid star requires a relatively small  $\Delta\varepsilon$, and a small  $\Delta\varepsilon$ do not support the existence of twin stars, thus the supermassive neutron star observation may also rule out the possibility of the existence of twin stars, which is consistent with what has been found in previous\ \cite{Christian2021}.
			\item
			A rough estimate of the lower limit of the dimensionless tidal deformability of neutron stars ranges from 2 to 3.
			It is shown that accurate observations on supermassive neutron stars with relatively large tidal deformability can provide effective constraints on the phase transition that softens EOS and its parameters $\Delta\varepsilon$, $\varepsilon_{tran}$ and $c_{sq}$. However, for supermassive neutron stars with small tidal deformability, even the third-generation gravitational detectors cannot reach the required precision.
			
		\end{enumerate}
		
		The phase transition may occur in other forms, such as hadron-quark crossover\ \cite{Blaschke2022,Huang2022,Huang2022K,Fujimoto2023}, which can stiffen the EOS in the phase transition region and bring different effects to hybrid stars. In future investigation, the macroscopic properties of supermassive hybrid stars can be further exploited to constrain the parameters of the crossover phase.

		Supermassive neutron stars contain both the extreme properties of neutron stars and the extreme properties of dense matter. If the observations confirm the existence of supermassive neutron stars, it will open a new window to enrich the theory of nuclear physics, gravity theory and astrophysics.

		\section{acknowledgement}
		This work is supported by NSFC (Grant No. 11975101) and Guangdong Natural Science Foundation (Grants No. 2022A1515011552 and No. 2020A151501820).
		%%%%²Î¿¼ÎÄÏ×

	\end{document}